# SOFT X-RAY EMISSIONS FROM PLANETS, MOONS, AND COMETS

A. Bhardwaj[1], G. R. Gladstone[2], R. F. Elsner[3], J. H. Waite, Jr.[4], D. Grodent[5], T. E. Cravens[6], R. R. Howell[7], A. E. Metzger[8], N. Ostgaard[9], A. N. Maurellis[10], R. E. Johnson[11], M. C. Weisskopf[3], T. Majeed[4], P. G. Ford[12], A. F. Tennant[3], J. T. Clarke[13], W. S. Lewis[2], K. C. Hurley[9], F. J. Crary[4], E. D. Feigelson[14], G. P. Garmire[14], D. T. Young[4], M. K. Dougherty[15], S. A. Espinosa[16], J.-M. Jahn[2]

[1]Space Physics Laboratory, Vikram Sarabhai Space Centre, Trivandrum 695022, India, spl_vssc@vssc.org
[2]Southwest Research Institute, San Antonio, TX 78228, USA
[3]NASA Marshall Space Flight Center, Huntsville, AL 35812, USA
[4]University of Michigan, Ann Arbor, MI 48109, USA
[5]LPAP, University of Liege, Belgium
[6]University of Kansas, Lawrence, KS 66045, USA
[7]University of Wyoming, Laramie, WY 82071, USA
[8]Jet Propulsion Laboratory, Pasadena, CA 91109, USA
[9]University of California at Berkeley, Berkeley, CA 94720, USA
[10]SRON National Institute for Space Research, Utrecht, The Netherlands
[11]University of Virginia, Charlottesville, VA 22903, USA
[12]Massachusetts Institute of Technology, Cambridge, MA 02139, USA
[13]Boston University, Boston, MA 02215, USA
[14]Pennsylvania State University, State College, PA 16802, USA
[15]Blackett Laboratory, Imperial College, London, England
[16]Max-Planck-Institut für Aeronomie, Katlenburg-Lindau, Germany

**ABSTRACT**

A wide variety of solar system bodies are now known to radiate in the soft x-ray energy (<5 keV) regime. These include planets (Earth, Jupiter, Venus, Saturn, Mars): bodies having thick atmospheres, with or without intrinsic magnetic field; planetary satellites (Moon, Io, Europa, Ganymede): bodies with thin or no atmospheres; and comets and Io plasma torus: bodies having extended tenuous atmospheres. Several different mechanisms have been proposed to explain the generation of soft x-rays from these objects, whereas in the hard x-ray energy range (>10 keV) x-rays mainly result from the electron bremsstrahlung process. In this paper we present a brief review of the x-ray observations on each of the planetary bodies and discuss their characteristics and proposed source mechanisms.

## 1. INTRODUCTION

The usually-defined range of x-ray photons spans 0.1-100 keV. Of this wide energy extent the soft x-ray energy band (<5 keV) is an important spectral regime for planetary remote sensing, as a large number of solar system objects are now known to shine at these wavelengths. These include Earth, Moon, Jupiter, Saturn, comets, Venus, Galilean satellites, Mars, Io plasma torus, and (of course) the Sun.

Since Earth's thick atmosphere efficiently absorbs x-ray radiation at lower altitudes (<30 km, even for hard x-rays), x-rays can only be observed from space by high-altitude balloon-, rocket-, and satellite-based instruments. But to observe most of the soft x-ray band one has to be above ~100 km from Earth's surface.

Terrestrial x-rays were discovered in the 1950s. The launch of the first x-ray satellite UHURU in 1970 marked the beginning of satellite-based x-ray astronomy. Subsequently launched x-ray observatories - Einstein, and particularly Rontgensatellit (ROSAT), made important contributions to planetary x-ray studies. With the advent of the latest and most sophisticated x-ray observatories – Chandra and XMM-Newton – the field of planetary x-ray astronomy is advancing at a much faster pace.

Earth and Jupiter, as magnetic planets, are observed to emanate strong x-ray emissions from their auroral (polar) regions, thus providing vital information on the nature of precipitating particles and their energization processes in planetary magnetospheres [1,2,3]. X-rays from low latitudes have also been observed on these planets. Saturn should also produce x-rays in the same way as Jupiter, although the intensity is expected to be weaker. Lunar x-rays have been observed from the sunlit hemisphere; and a small number of x-rays are also seen from the Moon's nightside [4]. Cometary x-rays are now a well-established phenomena; more than a dozen comets have been observed at soft x-ray energies [5,6].

The Chandra X-ray Observatory (CXO) has recently captured soft x-rays from Venus [7,8]. Martian x-rays are expected to be similar to those on Venus. More recently, using CXO [9] have discovered soft x-rays from the inner moons of Jupiter - Io, Europa, and probably Ganymede. The Io Plasma Torus (IPT) was also discovered recently by CXO to be a source of soft x-rays [9]. Though the x-rays from Jupiter were discovered in 1979 by Einstein observatory [cf. Ref. 2], the recent high spatial resolution observations by CXO/HRC-I have revealed a mysterious pulsating (period ~45 minutes) x-ray hot spot in the northern polar regions of Jupiter that have called into question our understanding of Jovian auroral x-rays [10].

In this paper we will present an overview of soft x-ray observations on planets, comets, and moons, and discuss the proposed emission production mechanisms. The Sun and heliosphere are the other known sources of soft x-rays in our solar system. The solar x-rays arise in the solar corona (which has a temperature of ~$10^6$ K), and consist of both line and continuum x-ray radiation that are produced by excitation of highly charged ions and thermal bremsstrahlung processes, respectively [11]. The heliospheric x-rays are observed as a part of the soft x-ray background [12,13], and are largely produced through charge transfer collision between highly stripped heavy solar wind ions and interstellar neutrals in the heliosphere [14]. The solar and heliospheric x-rays will not be discussed further as they are not covered by the topic of this paper.

## 2. EARTH

### 2.1. Auroral Emissions

Precipitating particles deposit their energy into the Earth's atmosphere by ionization, excitation, dissociation, and heating of the neutral gas. High-energy electrons or ions impacting the nucleus of atoms or molecules can lead to an emission of an x-ray photon by bremsstrahlung with an energy comparable to the energy of the incident particle. The x-ray bremsstrahlung production efficiency is proportional to $1/m^2$, where m is the mass of the precipitating particle. This implies that electrons are $10^6$ times more efficient than protons at producing x-ray bremsstrahlung. The production efficiency is a non-linear function of energy, with increasing efficiency for increasing incident energies. For example, for a 200 keV electron the probability of producing an x-ray photon at any energy below 200 keV is 0.5%, while the probability for a 20 keV electron to produce an x-ray photon below 20 keV is only 0.0057% [15].

The main x-ray production mechanism in Earth's auroral zones is electron bremsstrahlung, and therefore the x-ray spectrum of the aurora has been found to be very useful in studying the characteristics of energetic electron precipitation [16,17,18]. Since the x-ray measurements are not contaminated by sunlight, the remote sensing of x-rays can be used to study energetic electron precipitation on the nightside as well as on the dayside of the Earth [19]. Characteristic line emissions for the main species of the Earth's atmosphere, Nitrogen ($K_\alpha$ at 0.393 keV), Oxygen ($K_\alpha$ at 0.524 keV) and Argon ($K_\alpha$ at 2.958 keV, $K_\beta$ at 3.191 keV) will also be produced by both electrons and protons, but so far no x-ray observations have been made at energies where these lines are dominant compared to the x-ray bremsstrahlung.

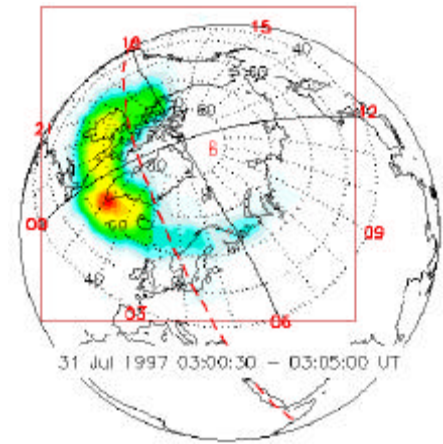

Fig.1. Auroral x-ray image of the Earth from the Polar PIXIE instrument (energy range 2.81-9.85 keV) obtained on July 31, 1997. The red box denotes the PIXIE field-of-view. The red dashed line and solid black line represent the day/night boundary and local noon, respectively. The grid in the picture is in geomagnetic coordinates and the numbers shown in red are magnetic local time.

X-ray photons from bremsstrahlung are emitted dominantly in the direction of the precipitating electron velocity. Consequently, the majority of the x-ray photons in Earth's aurora are directed towards the planet. These downward propagating x-rays, therefore, cause additional ionization and excitation in the atmosphere below the altitude where the precipitating particles have their peak energy deposition [e.g., Ref. 20,21]. The fraction of the x-ray emission that is moving away from the ground can be studied using satellite-based imagers, e.g, AXIS on UARS and PIXIE on POLAR.

Auroral x-ray bremsstrahlung has been observed from balloons and rockets since the 1960s and from spacecraft since the 1970s [22,23,24,25,16,17]. Due to the detector techniques that have been used, only x-rays above ~3 keV have been observed from the Earth's ionosphere. The PIXIE (Polar Ionospheric X-Ray Imaging Experiment) aboard Polar [26] is the first x-ray detector that provides true 2-D global x-ray image at energies >~3 keV (cf. Fig. 1). Because of the high apogee of the Polar satellite (~9 $R_E$), PIXIE is able to image the entire auroral oval with a spatial resolution of ~700 km for long duration when the satellite is around apogee. This has helped to study the morphology of the x-ray aurora and its spatial and temporal variation, and consequently the evolution of energetic electron precipitation during magnetic storms (days) and substorms (1-2 hours). Data from the PIXIE camera showed that the x-ray substorm brightens up in the midnight sector and has a prolonged and delayed maximum in the morning sector due to the scattering of drifting electrons [27]. Statistically the x-ray bremsstrahlung intensity peaks in the midnight substorm onset, is significant in the morning sector, and has a minimum in the early dusk sector [28]. During the onset/expansion phase of a typical substorm the electron energy deposition power is 60-90 GW, which produces 10-30 MW of bremsstrahlung x-rays [29]. Combining the results of PIXIE with the UV imager aboard Polar, it is possible to derive the energy distribution of precipitating electrons in the 0.1-100 keV range [18].

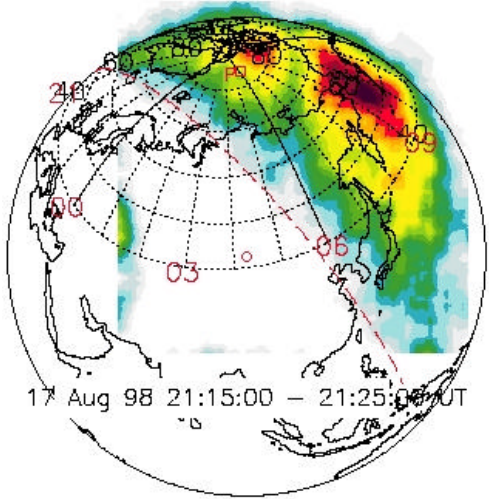

Fig. 2. X-ray image of Earth from the Polar PIXIE instrument for energy range 2.9-10.1 keV obtained on August 17, 1998, showing the dayside x-rays during a solar x-ray flare. The grid in the picture is in geomagnetic coordinates, and the numbers shown in red are magnetic local time. The terminator at the surface of the Earth is shown as a red dashed line.

### 2.2. Non-Auroral Emissions

The non-auroral x-ray background above 2 keV from the Earth is almost completely negligible except for brief periods during major solar flares [28]. However, at energies below 2 keV soft x-rays from the sunlit Earth's atmosphere have been observed even during quite (non-flaring) Sun conditions [e.g., ref. 30,31]. The two primary mechanisms for the production of x-rays from the sunlit atmosphere are: 1) the Thomson (coherent) scattering of solar x-rays from the electrons in the atomic and molecular constituents of the atmosphere, and 2) the absorption of incident solar x-rays followed by the emission of characteristic K lines of Nitrogen, Oxygen, and Argon. Fig. 2 shows the PIXIE image of Earth demonstrating the x-rays (2.9-10 keV) production in the sunlit atmosphere during a solar flare of August 17, 1998. During flares, solar x-rays light up the sunlit side of the Earth by Thomson scattering, as well as by fluorescence of atmospheric Ar to produce characteristic x-rays at 3 keV, which can be observed by PIXIE camera. The x-ray brightness can be comparable to that of a moderate aurora. Ref. [28] examined the x-ray spectra from PIXIE for two solar flare events during 1998. They showed that the shape of the measured x-ray spectra was in fairly good agreement with modeled spectra of solar x-rays subject to Thomson scattering and argon fluorescence in the Earth's atmosphere.

## 3. JUPITER

### 3.1. Auroral Emissions

The first detection of x-ray emissions from Jupiter was made by the satellite-based Einstein observatory in 1979 [32]. The emissions were detected in the 0.2-3.0 keV energy range from both poles of Jupiter. Analogous to the processes on Earth, it was expected that Jupiter's x-rays might originate as bremsstrahlung by precipitating electrons [33]. However, the power requirement for producing the observed emission with this mechanism ($10^{15}$-$10^{16}$ W) is more than two orders of magnitude larger than the input auroral power available as derived from Voyager and IUE observations of the ultraviolet aurora [cf. Ref. 2]. Ref. [32] suggested a mechanism implying K-shell line emissions from precipitating energetic sulfur and oxygen ions from the inner magnetosphere, with energies in the 0.3-4.0 MeV/nucleon range. The heavy ions are thought to emit x-rays by first charge stripping to a highly ionized state, followed by charge exchange and excitation through collisions with $H_2$. The bremsstrahlung process was further ruled out by theoretical models [34,35] showing that primary and secondary precipitating electrons in the 10-100 keV energy range are inefficient at producing the observed x-ray emissions.

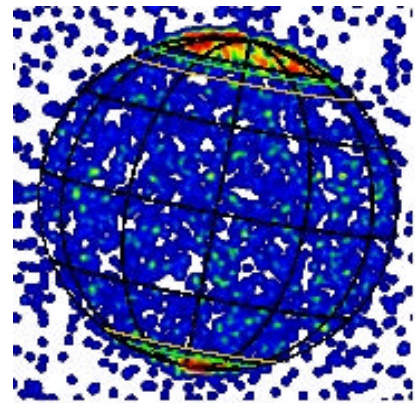

Fig. 3. Chandra x-ray image of Jupiter on 18 December 2000 generated from 10 hours of continuous observations. A jovicentric graticule with $30^0$ intervals is overplotted, along with the L=5.9 (orange lines) and L=30 (green lines) footprints of the magnetic field model. The image shows strong auroral x-ray emissions from high latitudes and rather uniform emissions from the disk. [from Ref. 10].

Furthermore, during its Jovian flyby, the Ulysses spacecraft did not detect significant emissions in the 27-48 keV energy range [36] as would have been the case if electron bremsstrahlung was a major process. Observations of Jupiter x-ray's emissions by ROSAT [37] supported the suggestion of [32] and the model calculations [34,35] that precipitating energetic (>700 keV per nucleon) S and O ions are most probably responsible for the x-ray emissions from Jupiter. A detailed modeling of the x-ray production [38,39] suggests that recombination lines from highly charged precipitating O and S ions mainly contribute to the soft x-rays detected by ROSAT.

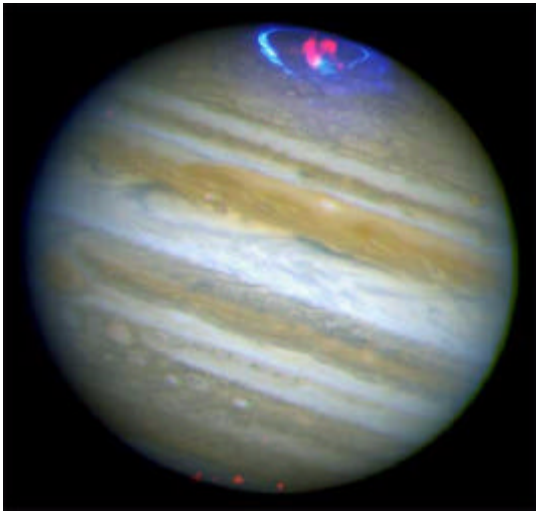

Fig. 4. This composite image displays x-ray data from Chandra (magenta) and ultraviolet data from Hubble Space Telescope (blue) overlaid on an optical image of Jupiter. While Chandra observed Jupiter for an entire 10-hour rotation period on December 18, 2000, this image shows a 'snapshot' of a single 45-minute X-ray pulse. [from http://chandra.harvard.edu/photo/2002/0001/0001_xray_opt_uv.jpeg].

Recent high-spatial resolution observations of Jupiter with the Chandra telescope [10] reveal that most of Jupiter's northern auroral x-rays come from a "hot spot" located significantly poleward of the latitudes connected to the inner magnetosphere (cf. Figs. 3, 4). The hot spot is fixed in magnetic latitude and longitude and occurs in a region where anomalous infrared and ultraviolet emissions (the so-called "flares") have also been observed. Its location must connect along magnetic field lines to regions in the Jovian magnetosphere well in excess of 30 Jovian radii from the planet (cf. Fig. 3), a region where there are insufficient S and O ions to account for the hot spot [10]. More surprising, the hot spot x-rays pulsate with an approximately 45-min period, similar to that reported for high-latitude radio and energetic electron bursts observed by near-Jupiter spacecraft [cf. Ref. 10].

These results invalidate the idea that Jovian auroral x-ray emissions are mainly excited by steady precipitation of heavy energetic ions from the inner magnetosphere [cf. Ref. 2]. Instead, the x-rays seem to result from currently unexplained processes in the outer magnetosphere that produce highly localized and highly variable emissions over an extremely wide range of wavelengths. In any case, the power needed to produce the brightest ultraviolet "flares" seen in the same polar cap region as the x-ray hot spot is a few tens of TW, much less than the estimated power of a few PW needed to produce the observed x-rays by electron bremsstrahlung. Thus, electron bremsstrahlung still seems to fail in explaining the observed Jovian x-rays hot spot. One possible source of Jovian x-rays production is via charge-exchange of solar wind ions that penetrate down the atmosphere in the magnetic cusp region. But in this case the solar wind ions would have to be accelerated to much higher energies (100s of keV), probably by parallel electric field or wave particle interactions, to generate sufficient luminosity to account for the observations. In summary, at the present time the origin of the Jovian x-rays and its source is still an open issue.

### 3.2. <u>Non-Auroral Emissions</u>

Soft x-ray emissions with brightnesses of about 0.01-0.2 Rayleighs were observed from the equatorial regions of Jupiter using the ROSAT/HRI. It was proposed [40] that the equatorial emission, like the auroral emission, may be largely due to the precipitation of energetic (>300 keV/amu) sulfur or oxygen ions into the atmosphere from the radiation belts. Further evidence for a correlation between regions of low magnetic field strength and enhanced emission [41] lent additional support to this mechanism, since it can be assumed that the loss cone for precipitating particles is wider in regions of weak surface magnetic field. However, [42] showed that two alternative mechanisms should not be overlooked in the search for a complete explanation of low-latitude x-ray emission, namely elastic scattering of solar x-rays by atmospheric neutrals and fluorescent scattering of carbon K-shell x-rays from methane molecules located below the Jovian homopause. Modeled brightnesses agree, up to a factor of two, with the bulk of low-latitude ROSAT/PSPC measurements which suggests that solar photon scattering (~90% elastic scattering) may act in conjunction with energetic heavy ion precipitation to generate Jovian non-auroral x-ray emission. The solar x-ray scattering mechanism is also suggested from the correlations of Jovian emissions with the F10.7 solar flux and of the x-ray limb with the bright visible limb [41]. During the December 2000 observations by Chandra HRC-I, the disk-averaged emitted x-ray power was about 2 GW [10], but the signal-to-noise ratio of the disk emission was not adequate to show the limb brightening expected by solar photon-driven x-ray emission.

## 4. MOON

Though it is the Earth's nearest planetary body, the Moon has been relatively little studied at x-ray wavelengths. Other than the discovery observation by [4] using the ROSAT PSPC and a detection by Advanced Satellite for Cosmology and Astrophysics (ASCA) [43], most recent high-energy remote sensing of the Moon has been made at extreme- and far-ultraviolet wavelengths [e.g., Ref. 44,45,46]. However, as noted by [47], x-ray fluorescence studies could provide an excellent way to determine the elemental composition of the lunar surface by remote sensing, since the soft x-ray optical properties of the lunar surface should be dominated by elemental abundances (rather than mineral abundances, which determine the optical properties at visible and longer wavelengths). Although reflection of the strong solar lines likely dominates the soft x-ray spectrum of the Moon, the detection of weaker emissions due to L- and M-shell fluorescence would provide a direct measure of specific elemental abundances.

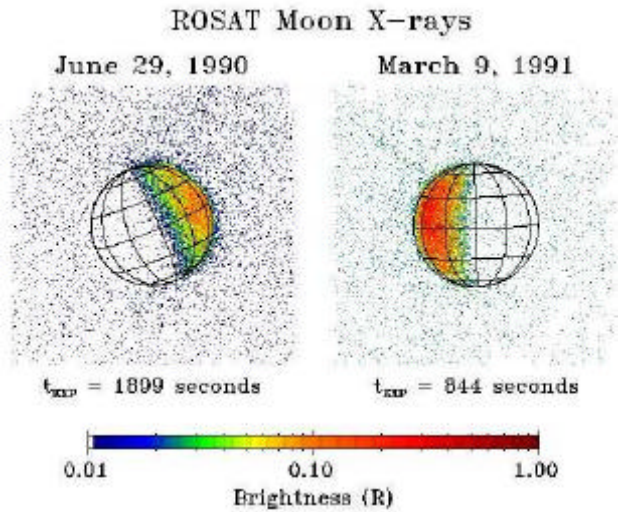

Fig. 5. ROSAT soft x-ray (0.1-2 keV) images of the Moon at first (left side) and last (right side) quarter. The dayside lunar emissions are thought to be primarily reflected and fluoresced sunlight, while the origin of the faint but distinct nightside emissions is uncertain. The brightness scale in R assumes an average effective area of 100 cm$^2$ for the ROSAT PSPC over the lunar spectrum.

Fig. 5 shows ROSAT data; the left image shows the [4] data, while the right image is unpublished data from a lunar occultation of the bright x-ray source GX5-1 (the higher energy x-rays from GX5-1 have been suppressed in this figure, but a faint trail to the upper left of the Moon remains). The power of the reflected and fluoresced x-rays observed by ROSAT in the 0.1-2 keV range coming from the sunlit surface was determined by [4] to be only 73 kW, making the Moon the faintest x-ray source in the sky (the flux measured was $2.5\times10^{-12}$ erg cm$^{-2}$ s$^{-1}$).

While the dayside lunar soft x-rays are reflected and fluoresced sunlight, the faint but distinct lunar nightside emissions are a matter of controversy. Ref. [4] suggested that solar wind electrons of several hundred eV might be able to impact the night side of the Moon on the leading hemisphere of the Earth-Moon orbit around the sun. However, this was before the GX5-1 data were acquired, which clearly show lunar nightside x-rays from the trailing hemisphere as well. Another possible explanation is the accepted mechanism for comet x-rays, heavy ion solar wind charge exchange (SWCX) [e.g., Ref. 6]. In this case, however, the heavy ions in the solar wind would be charge exchanging with geocoronal H atoms that lie between the Earth and Moon but lie outside the Earth's magnetosphere.

Future observations of the Moon's x-rays by Chandra and XMM are likely, and there is a planned x-ray spectrometer D-CIXS on ESA's SMART-1 lunar mission [48] that will provide global coverage of ambient x-ray emission. This will greatly improve upon the elemental abundance maps produced by the x-ray spectrometers on Apollo 15 and 16 [49].

## 5. COMETS

X-ray emission from a comet was first discovered in 1996 with the ROSAT observations of comet Hyakutake [50]. Extreme ultraviolet (EUV) emission was also detected from this comet by the EUVE satellite [51]. Since the initial discovery of cometary x-ray emission, it has been shown that active comets are almost always EUV and soft x-ray sources [52]. A thorough review of this topic has just appeared [6], so only a brief summary is provided here.

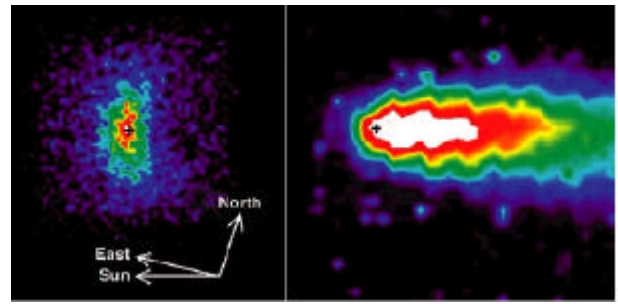

Fig. 6. X-ray and visible images of comet C/LINEAR 1999 S4. Left: Chandra x-ray (0.2-0.8 keV) July 14, 2002 ACIS-S image. Right: Visible light image of comet taken on July 14, 2002 showing a symmetric coma and a long anti-solar tail. The plus sign mark the position of the nucleus. [from Ref. 5].

The key observational features of cometary x-ray emission (cf. Fig. 6) are now summarized. The x-ray emission is "very soft" with photon energies of a few hundred eV or less. High-resolution x-ray spectra of comets C/Linear 1999 S4 [5] and McNaught-Hartley (C/1999 T1) [53] have recently been measured by the CXO. Emission lines associated with highly charged ions (in particular, $O^{6+}$) are evident in these spectra. A spectrum of comet Hyakutake from the EUVE satellite [54] also displays line emission. Cometary x-ray luminosities are quite large and tend to correlate with the gas production rate [e.g., Ref. 55]. The comet Hyakutake x-ray luminosity measured by ROSAT was about 1 GW. Cometary x-ray emission is spatially very extensive with observed emission out to radial distances from the cometary nucleus of $10^5$-$10^6$ km [e.g., Ref. 5,50] The emission is also time variable and has been shown to correlate with the solar wind flux [e.g., Ref. 56].

Several cometary x-ray emission mechanisms were proposed following the initial discovery. These include bremsstrahlung associated with solar wind electron collisions with cometary neutrals and ions, K-shell ionization of neutrals by electron impact, scattering of solar photons by cometary dust, and charge transfer of solar wind heavy ions with neutrals [cf. Ref. 6]. The SWCX mechanism [57] has gradually won favor. In this mechanism highly-charged solar wind heavy ions (e.g., $O^{6+}$, $O^{7+}$, $C^{6+}$, $Ne^{9+}$,...) undergo charge transfer collisions when they encounter cometary neutrals. The product ions are invariably left in highly excited states and emit EUV and soft x-ray photons. This mechanism is able to explain the luminosity, spatial distribution [58], time variability [56], and spectrum [5,53,59,60] of the x-ray emission.

## 6. VENUS

In January 2001 the CXO discovered soft x-ray emissions from Venus [7,8]. Observations were performed with the CXO's ACIS-I high-resolution imaging camera (cf. Fig. 7) and with the LETG/ACIS-S high-resolution grating spectrometer. The x-ray emission originated from the sunlit hemisphere, exhibited noticeable limb brightening (when compared to visible disk), and consisted primarily of O-$K_\alpha$ at ~530 eV. The C-$K_\alpha$ line emission at ~280 eV was also detected, with marginal evidence for N-$K_\alpha$ emission at ~400 eV. The carbon emission might also include a 290 eV line from $CO_2$ and CO. The derived energy fluxes in the oxygen and carbon lines were ~20× $10^{-14}$ erg $cm^{-2}$ $s^{-1}$ for O-$K_\alpha$ and ~5×$10^{-14}$ erg $cm^{-2}$ $s^{-1}$ for C-$K_\alpha$. The total power in Venusian x-rays was estimated to be 30-70 MW. The Chandra data also indicated possible time variability on a timescale of a few minutes.

The observers [8] argue persuasively that the x-ray emission from Venus results from fluorescent scattering of solar x-rays, as predicted by [61], and not from charge exchange between heavy ions in the solar wind and neutral atoms in the Venusian atmosphere. Their detailed modeling of the interaction between solar x-rays and the planetary atmosphere showed that the fluorescent scattering occurred ~110 km or higher above the planet's surface. The amount of limb brightening predicted by their models depended sensitively on the chemical composition and the density profile in Venus's upper atmosphere.

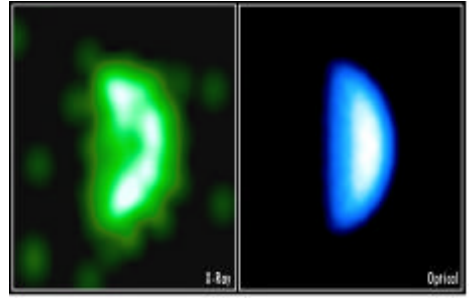

Fig. 7. X-ray image of Venus obtained with Chandra ACIS-I on 13 January 2001. Right is an optical image of Venus. [from Ref. 8].

## 7. GALILEAN SATELLITES

Recently the CXO has discovered [9] x-ray emission from the Galilean satellites (cf. Fig. 8). The CXO observations of the Jovian system were made on 25-26 November 1999 for 86.4 ks with the ACIS-S instrument and on 18 December 2000 for 36.0 ks with the HRC-I instruments. The time tagged nature of the CXO data makes it possible to correct for varying satellite motions, and with ACIS it is also possible to filter the data by energy for optimum sensitivity. During the ACIS-S observation, Io and Europa were detected with a high degree of confidence, and Ganymede at a lesser degree of confidence. Io was also detected with high confidence during the shorter HRC-I observation. Over the nominal energies of 300-1890 eV range detected by ACIS-S, the x-ray events show a clustering between 500 and 700 eV, probably dominated by the oxygen $K_\alpha$ line at 525 eV. The estimated energy fluxes at the telescope and power emitted are $4\times10^{-16}$ erg $cm^{-2}$ $s^{-1}$ and 2.0 MW for Io, and $3\times10^{-16}$ erg $cm^{-2}$ $s^{-1}$ and 1.5 MW for Europa. Ganymede was roughly a third as luminous as Io. Callisto was not detected in either set of data.

The most plausible emission mechanism is inner (K shell) ionization of the surface (and perhaps incoming magnetospheric) atoms followed by prompt x-ray emission. Oxygen should be the dominant emitting atom in either a silicate or $SO_2$ surface (Io) or in an icy one (the outer Galilean satellites). It is also the most common heavy ion in the Jovian magnetosphere. The extremely tenuous atmospheres of the satellites are transparent to x-ray photons with these energies, and also to much of the energy range of the incoming ions. However, oxygen absorption of the 525 eV line is such that the x-rays must originate in the top ~10 microns of the surface in order to escape. Simple estimates suggest that excitation by incoming ions dominates over electrons and that the x-ray flux produced is sufficient to account for the observations.

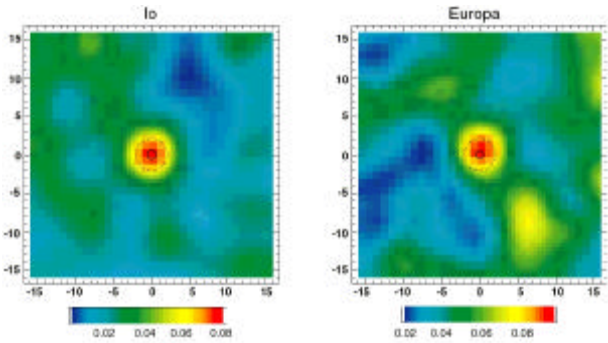

Fig. 8. Chandra ACIS-I image (0.2-2 keV) of Io and Europa obtained on November 25-26, 1999. The solid circle shows the size of the satellite (the radii of Io and Europa are 1821 km and 1560 km, respectively), and the dotted circle the size of the detect cell. The axes are labeled in arcsec (1 arcsec ≈ 2995 km) and the scale bar is in units of smoothed counts per image pixel (0.492 by 0.492 arcsec). [from Ref. 9].

Detailed models are required for verifying this picture and also for predicting the strengths of $K_\alpha$ lines for elements other than oxygen, especially heavier ones such as Na, Mg, Al, Si, and S. Within this framework, it is possible to constrain the surface composition of these moons from x-ray observations, but this requires a greater signal-to-noise ratio than provided by the Chandra observations. The detection of x-ray emission from the Galilean satellites thus provides a direct measure of the interactions of magnetosphere of Jupiter with the satellite surfaces.

## 8. IO PLASMA TORUS

The Io Plasma Torus (IPT) is known to emit at EUV energies and below [62,63,64], but it was a surprise when CXO discovered that it was also a soft x-ray source [9]. The Jovian system has so far been observed with Chandra using the ACIS-S high-spatial-resolution imaging camera, which also has modest energy resolution, for two Jovian rotations in November 1999, and using the HRC-I high-spatial-resolution camera, with essentially no energy resolution, for one rotation in December 2000. X-ray emission from the IPT is present in both observations. The ACIS-S spectrum was consistent with a steep power-law continuum (photon index 6.8) plus a gaussian line (complex) centered at ~569 eV, consistent with $K_\alpha$ emission from various charge species of oxygen. Essentially no x-rays were observed above this spectral feature, consistent with the steepness of the power-law continuum. The 250-1000 eV energy flux at the telescope aperture was $2.4\times10^{-14}$ erg cm$^{-2}$ s$^{-1}$, corresponding to a luminosity of 0.12 GW, and was approximately evenly divided between the dawn and dusk side of Jupiter.

However, the line emission originated predominantly on the dawn side. During the ACIS-S observation (cf. Fig. 9), Io, Europa, and Ganymede were on the dawn side, while Callisto was on the dusk side. For the HRC-I observation, the x-ray emission was stronger on the dusk side, approximately twice that observed on the dawn side. During the HRC-I observation, Io, Europa, and Ganymede were on the dusk side, while Callisto was on the dawn side.

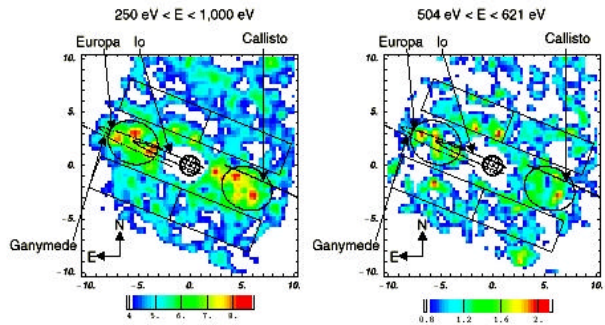

Fig. 9. Chandra ACIS-I image of Io plasma torus obtained on November 25-26, 1999. The axes are labelled in units of Jupiter's radius, $R_J$ and the scale bar is in units of smoothed counts per image pixel (7.38 by 7.38 arcsec). For this observation, Jupiter's radius corresponds to 23.8 arcsec. The paths traced by Io (solid line to the east), Europa (dashed line), Ganymede (dotted line), and Callisto (solid line to the west) are marked on the image. The regions bounded by rectangles were used to determine background. The regions bounded by ellipses were defined as x-ray source regions. [from Ref. 9].

The physical origin of the x-ray emission from the IPT is not yet fully understood. According to the estimates given in [9], fluorescent x-ray emission excited by solar x-rays, even during flares from the active Sun, charge-exchange processes, previously invoked to explain Jupiter's x-ray aurora [e.g., Ref. 2,38] and cometary x-ray emission [e.g., Ref. 6,57], and ion stripping by dust grains fail to account for the observed emission. Assuming bremsstrahlung emission of soft x-rays by non-thermal electrons in the few hundred to few thousand eV range, with a kappa, or generalized Lorentzian, distribution with a temperature of 10 eV and an index of $\kappa = 2.4$ [65], which is consistent with the in-situ Ulysses observations, [9] estimated an IPT soft x-ray luminosity of 0.03 GW. This falls short of but is a significant fraction of the observed luminosity of 0.12 GW.

## 9. MARS

There seems to be a tentative detection of Martian x-rays in the ROSAT data [66]. However, so far no x-ray emission has been unambiguously detected from Mars, but this situation is expected to change soon. As at Venus, absorption of solar x-rays in either the carbon or oxygen K-shells followed by fluorescent emission of x-rays is suggested as the dominant process of x-rays production on Mars [61]. The predicted total soft x-ray intensity is 0.15 R, corresponding to an x-ray luminosity of about 2.5 MW [61], which exceeds the x-ray luminosity expected from the solar wind charge exchange mechanism on Mars [67]. A simulated image of SWCX-produced x-ray emission at Mars [67] indicates that this emission has a very different spatial morphology than the fluoresced x-ray emission [61], and future observations must be able to distinguish between the two processes.

## 10. SATURN

The existence of a magnetosphere at Saturn and the presence within it of energetic electrons and ions, both first observed by instruments on Pioneer 11 and Voyagers 1 and 2, provide the conditions under which auroral emission can be expected. The first indication of an aurora at Saturn came from measurements in the ultraviolet by Pioneer 11, followed up by observations with the IUE satellite, and confirmed by the Voyager 1 and 2 flyby encounters, whose results included localizing the sources of emission to regions near the pole in both hemispheres [e.g., Ref. 68; cf. Ref. 2 for review].

These particle and field properties make it probable that auroral x-ray emission occurs at Saturn. The UV emission can be accounted for by electrons precipitating along high latitude field lines into Saturn's atmosphere [2]. X-ray emission could be generated by bremsstrahlung involving the high-energy portion of the precipitating electron distribution. Alternatively, energetic (~1 MeV C, N, and O) ions have been observed in Saturn's magnetosphere [69]. A fraction of these could precipitate into Saturn's atmosphere at high latitudes, generating x-rays through charge exchange reactions as has been postulated to occur at Jupiter [2]. It has also been suggested that a potential source of heavy ions, additional to solar wind injected plasma, is nitrogen escaping from the atmosphere of Saturn's satellite Titan [70]. Other possible contributory sources, particularly in view of the recent observation of x-ray emission from three of the four Galilean satellites of Jupiter [9], are Saturn's rings and satellite surfaces, to the degree these are exposed to bombardment by energetic particle fluxes.

Following the discovery of x-ray emission from Jupiter with the Einstein Observatory [32], an observation of Saturn was undertaken [71]. No x-ray emission was seen. With the conversion of count rate into flux dependent on spectral shape, and spectral shape a consequence of the production mechanism, the 3σ upper limit at Earth from this observation was calculated to be $5 \times 10^{-13}$ erg cm$^{-2}$ s$^{-1}$ if the mechanism is dominated by ion-produced characteristic lines, and $2 \times 10^{-13}$ erg cm$^{-2}$ s$^{-1}$ if the mechanism is electron bremsstrahlung. Assuming the latter, and basing the expected x-ray flux on the intensity of the UV aurora observed by Voyager 2, a model calculation of the expected energy flux in the 0.2-3.0 keV energy range yielded a value of $8\times10^{-16}$ erg cm$^{-2}$ s$^{-1}$, more than two orders of magnitude below the observed upper limit.

In 1992, [72] observed Saturn with the ROSAT position sensitive proportional counter, an instrument with superior sensitivity to soft x-rays relative to the corresponding instrument on the Einstein Observatory. The observed events were grouped into two energy bands. The harder band saw nothing significant, but the soft band, ranging from ~0.1 to 0.55 keV, recorded almost three times as many counts as expected from background. The corresponding net energy flux is $1.9 \times 10^{-14}$ erg cm$^{-2}$ s$^{-1}$. This is more than an order of magnitude greater than the model estimate for x-ray production based on electron bremsstrahlung [71], an estimate more likely to be on the high than the low side [73]. Furthermore, the observed flux amounts to 24% of the flux observed from Jupiter under identical instrument conditions after removing the effect due to their different distances from Earth. Even if the predominant mechanism at both planets is some form of ion precipitation, a mechanism which produces x-rays more efficiently than electron bremsstrahlung, but which recent Chandra Observatory results have called into question at Jupiter [10] and which seems less likely at Saturn, emission at Saturn would be expected to be less than 10% of that from Jupiter, a ratio more consistent with that of the observed auroral intensities [74]. The non-auroral x-rays from Saturn is expected to have a predominantly solar-driven elastic scattering, much like Jupiter (predominantly $H_2$ scattering with a small $CH_4$ component) with perhaps about a third as much integrated non-auroral power as Jupiter. Additional observations of Saturn with the Chandra or XMM-Newton Observatories are called for to follow up on the intriguing result from ROSAT.

## 11. SUMMARY

Table 1 summaries our current knowledge of the x-ray emissions from the planetary bodies that have been observed to produce soft x-rays. Several other solar system bodies, which include Titan, Uranus, Neptune, and inner-icy satellites of Saturn, are also expected to be x-ray sources, but are yet to be detected.

The x-rays from solar system bodies are relatively weak (< few GW) and are from a much colder (T <$10^3$ K) environment than x-rays from stars and other cosmic bodies (>TW, T >$10^6$ K). Nonetheless, the studies of planetary x-rays help advance our understanding of basic plasma-neutral (in both gas and solid phase) interactions that are important within our solar system, and plausibly in the extra-solar planetary systems as well.

Table 1. Summary of the characteristics of soft x-rays from solar system bodies

| Object | Emitting Region | Power Emitted[a] | Special Characteristics | Possible Production Mechanism | References[b] |
|---|---|---|---|---|---|
| Earth | Auroral atmosphere | 10-30 MW | Correlated with magnetic storm and substorm activity | Bremsstrahlung from precipitating electrons | [29] |
| Earth | Non-auroral atmosphere | 40 MW | Correlated with solar x-ray flux | Scattering of solar x-rays by atmosphere | [31] |
| Jupiter | Auroral atmosphere | 0.4-1 GW | Pulsating (~45 min) x-ray hot spot in north polar region | Energetic ion precipitation from magnetosphere and/or solar wind + electron bremsstrahlung + ? | [32,10] |
| Jupiter | Non-auroral atmosphere | 0.5-2 GW | Quite uniform over disk | Resonant scattering of solar x-rays + Ion precipitation from ring current | [40,42,10] |
| Moon | Surface - Dayside<br>- Nightside | 0.07 MW | Nightside emissions are ~1% of the dayside emissions | Scattering of solar x-rays by the surface elements on dayside. Electron bremsstrahlung + SWCX of geocorona? | [4,43] |
| Comets | Sunward-side coma | 0.2-1 GW | Intensity peaks in sunward direction ~$10^5$-$10^6$ km ahead of cometary nucleus | SWCX with cometary neutrals + other mechanisms | [5,6,50] |
| Venus | Sunlit atmosphere | 50 MW | Emissions come from ~120-140 km above the surface | Fluorescent scattering of solar x-rays by C and O atoms in the atmosphere | [7,8,61] |
| Io | Surface | 2 MW | Emissions from upper few microns of the surface | Energetic Jovian magnetospheric ions impact on the surface + ? | [9] |
| Europa | Surface | 1.5 MW | Emissions from upper few microns of the surface | Energetic Jovian magnetospheric ions impact on the surface + ? | [9] |
| Io Plasma Torus | Plasma torus | 0.1 GW | Dawn-dusk asymmetry observed? | Electron bremsstrahlung + ? | [9] |
| Saturn | Auroral and non-auroral atmosphere | 0.4 GW | Plausibly similar to Jovian x-rays | Electron bremsstrahlung + scattering of solar x-rays | [72] |
| Mars | Atmosphere | 3 MW | Probable detection? | Solar fluorescence + SWCX | [66,61] |

[a]The values quoted are "typical" values at the time of observation. X-rays from all bodies are expected to vary with time. For comparison the total x-ray luminosity from the Sun is $10^{20}$ W and that from the heliosphere $10^{16}$ W [75].
[b]Only representative references are given.
SWCX ≡ Solar wind charge exchange = charge exchange of heavy highly ionized solar wind ions with neutrals.